\documentclass[12pt]{article}
\usepackage{amssymb}
\usepackage[T1]{fontenc}
\usepackage[utf8]{inputenc}
\usepackage[intlimits]{amsmath}
\usepackage{enumerate}
\usepackage[all]{xy}

\newtheorem{theorem}{Theorem}
\newtheorem{lemma}{Lemma}

\usepackage{multirow}

\title{Periodic one-point rank one commuting difference operators}
\author{Alina Dobrogowska\footnote{A.D. was supported by  the Polish Ministry of Science and Higher Education under subsidy for maintaining the research potential of the Faculty of Mathematics and Informatics, University of Bialystok}  \ and   Andrey E. Mironov\footnote{A.E.M. was supported by RFBR (grant 18-01-00411) and by the State Maintenance Program for the Leading Scientific Schools of the Russian Federation (grant NSh-5913.2018.1)}}
\date{}

\begin{document}

\sloppy
\maketitle

\begin{abstract}
In this paper we study one-point rank one commutative rings of difference operators. We find conditions on spectral data which  specify such operators with periodic coefficients.
\end{abstract}

\noindent
{Keywords: commuting difference operators}

\section{Introduction and Main Results}

In this paper we study one--point rank one commuting difference operators with periodic coefficients.

Let us consider a (maximal) commutative ring ${\mathcal A}$ of difference operators consisting of operators of the form
$$
 L_m= \sum_{i=-N_-}^{i=N_+}u_i(n)T^i, \quad u_{N_+}=1, N_+,N_-\geq 0,
$$
where $T$ is a shift operator, $T\psi(n)=\psi(n+1)$, $ m=N_++N_-$ is the oder of $L_m$  (assuming $u_{N_{-}}\neq 0$). The operator $L_m$ acts on the space of formal functions $\{\psi:\mathbb{Z}\longrightarrow \mathbb{C}\}$. The ring ${\mathcal A}$ is isomorphic to a ring of rational functions on spectral curve $\Gamma$ with poles in points $q_1,\dots,q_s\in \Gamma$ (see \cite{Kri}). Common eigenfunctions of operators from 
${\mathcal A}$ form a vector bundle of rank $l$ over $\Gamma\backslash\{ \cup_{j=1}^{s}q_j\}.$ More precisely, there is a vector-function $\psi(n,P)=(\psi_1(n,P),\dots,\psi_l(n,P)), P\in\Gamma\backslash\{ \cup_{j=1}^{s}q_j\}$ which is called {\it Baker--Akhiezer function}, such that every operator $L_m\in{\mathcal A}$ corresponds to a meromorphic function $f(P)$ on $\Gamma$ with poles in $q_1,\dots,q_s$
$$
 L_m\psi=f\psi.
$$
Moreover, $m=lm'$, where $m'$ is  the degree of the pole divisor of $f.$ The operators from ${\mathcal A}$ are called {\it $s$-point rank $l$ operators}.

 Two--point rank one operators were classified in \cite{Kri,Mum}. Baker--Akhiezer function of such operators can be reconstructed from Krichever's spectral data  \cite{Kri}. One-point rank $l>1$  operators were discovered by Krichever and Novikov in \cite{KriNov}. 
 Spectral data for one-point rank one operators were found in \cite{MauMir}. Those operators contain the shift operator $T$ only in positive power.  Recall that the spectral data for such operators has the form (see \cite{MauMir}) 
 $$
S=\{\Gamma, \gamma,  P_n, q,k^{-1}\}.
$$
Here  $\Gamma$ is a Riemannian surface of genus $g$ (we do not consider singular spectral curves), 
$\gamma=\gamma_1+\ldots+\gamma_g$ is a non-special divisor on $\Gamma$, 
$P_n\in\Gamma$, $n\in\mathbb{Z}$ is a set of general points, $q\in\Gamma$ is a fixed point, $k^{-1}$ is a local parameter near $q$.
There is a unique Baker--Akhezer function  $\psi(n,P),n\in\mathbb{Z}, P\in\Gamma$ which is rational function on $\Gamma$ and satisfies the following properties
\begin{itemize}
\item if $n>0,$ then the zero and pole divisor of $\psi$  has the form
$$(\psi(n,P))= \gamma(n)+P_0+ \ldots +P_{n-1}-\gamma-nq,$$
\item if $n<0,$ then the zero and pole divisor of $\psi$  has the form
 $$(\psi(n,P))= \gamma(n)-P_{-1}- \ldots -P_n-\gamma-nq,$$
\item if $n=0$ then $\psi(n,P)=1,$
\item in a neighborhood of $q$ the function $\psi$ has the following expansion 
$$
\psi(n,P)=k^n+O(k^{n-1}).
$$
\end{itemize}
Here $\gamma(n)=\gamma_1(n)+\ldots+\gamma_g(n),\ n\ne 0$ is some divisor on $\Gamma.$  Further we will use the following notation $\gamma(0)=\gamma.$
For arbitrary meromorphic function $f(P)$ on $\Gamma$ with the unique pole in $q$ of order $m$ there is a unique operator 
$$
 L_m=T^m+u_{m-1}(n)T^{m-1}+\ldots+u_0(n)$$  
such that $L_m\psi=f\psi$, see \cite{MauMir}.

If in the spectral data $S$ all points $P_n$ coincide, $P_n=q^+,$ then we get the two-point Krichever's construction \cite{Kri}.

 One-point rank one operators were studied in \cite{MauMir, MauMir1,  MauMir2}, in particular some explicit examples of such operators {were given. That class of operators is very interesting because, for example with the help of those operators one can construct a discretization of the Lam\'e op
 erator  preserving the spectral curve. More precisely, 
let $\wp (x)$, $\zeta (x)$ be the Weierstrass functions. We define the function $A_g(x,\varepsilon)$ by the following formulas
\begin{align}
& A_1=-2 \zeta (\varepsilon)-\zeta (x-\varepsilon)+ \zeta (x+\varepsilon),\nonumber\\
& A_2=-\frac{3}{2}\left( \zeta (\epsilon)+\zeta (3\varepsilon)+\zeta (x-2\varepsilon)-
\zeta (x+2\varepsilon)\right),\nonumber\\
& A_g=A_1
\prod_{i=1}^{g_1}\left(1+ \dfrac{\zeta (x-(2i+1)\varepsilon )-\zeta (x+(2i+1)\varepsilon)}
{\zeta (\varepsilon )+\zeta ((4i+1)\varepsilon )}\right),  \textrm{ for odd } g=2g_1+1,\nonumber\\
& A_g=A_2\prod_{i=2}^{g_1}\left(1+ \dfrac{\zeta (x-2i\varepsilon )-\zeta (x+2i\varepsilon)}
{\zeta (\varepsilon )+\zeta ((4i-1)\varepsilon )}\right),  \textrm{ for even } g=2g_1.\nonumber
\end{align}
The operator 
\begin{equation}\label{eq1}
L_2=\frac{1}{\varepsilon^2}T^2_{\varepsilon}+A_g(x,\varepsilon)\frac{1}{\varepsilon}T_{\varepsilon}+\wp (\varepsilon)
\end{equation}
commutes with the operator $L_{2g+1}$, operators $L_2,L_{2g+1}$ are rank one one-point operators. In the above formulas it is assumed that $T_{\varepsilon}\psi(x)=\psi(x+\varepsilon).$ The operator $L_2$ has the following expansion
$$
 L_2=\partial_x^2-g(g+1)\wp(x)+O(\varepsilon).
$$
For small $g$ it is checked that the spectral curve of the pair $L_2,L_{2g+1}$ coincides with the spectral curve of the Lam\'e ope\-rator $\partial_x^2-g(g+1)\wp(x)$, see \cite{MauMir1}. Probably this class of difference operators can be used for the construction of a discretization of arbitrary finite-gap one dimensional Schr\"odinger operators. Note that the operator (\ref{eq1}) is periodic. So, for the discretization of the finite-gap operators it is useful to find the condition when rank one one-point operators are periodic with real coefficients. This is the main motivation of this paper.

 In the next theorem we formulate periodicity and reality conditions of the coefficients of the operators.

\begin{theorem}
Coefficients of one-point rank one operators corresponding to the spectral data
$$
S=\{\Gamma, \gamma,  P_n, q,k^{-1} \}
$$
are $N$--periodic, $N\in{\mathbb N}$,  if and only if 
$$
  P_{n+N}=P_n,
$$
and there is a meromorphic function $\lambda(P)$ on $\Gamma$  with a divisor of zeros and poles of the form
$$
\left(\lambda(P)\right)=P_0+\ldots +P_{N-1}-Nq.
$$

 Let us assume that the  spectral curve  $\Gamma$ admits an antiholomorphic involution
$$ \tau : \Gamma \longrightarrow \Gamma, \quad \tau^2=id.$$
If 
\begin{equation}\label{eq4n}
\tau(P_n)=P_n, \quad  \tau(\gamma)=\gamma,\quad  \tau(q)=q, \quad \tau(k)=\overline{k},
\end{equation} 
then the Baker--Akhiezer function satisfies the identity
\begin{equation}\label{eq4}
 \psi(n,P)=\overline{\psi(n,\tau(P))},
\end{equation}
and  if additionally 
$$
 \tau(f(P))=\overline{f(P)},
$$
then the coefficients of the operator $L_m$ corresponding to the function $f(P),$ $L_m\psi=f\psi,$ are real.
\end{theorem}

  In the case of two-point rank one operators the analogue of Theorem 1 was proved in \cite{K}. In the two-point case we have 
 $(\lambda)=N q^+-N q.$

\section{Proof of Theorem 1}
In the beginning we prove the second part of the theorem.  The proof of this part is usual. The identity (\ref{eq4}) follows from the uniqueness of the Baker--Akhiezer function with the fixed spectral data. Indeed, from (\ref{eq4n}) it follows that the function $\overline{\psi(n,\tau(P))}$ satisfies the same conditions as $\psi(n,P),$ hence we get (\ref{eq4}).  

We have
$$
 L_m\psi(n,\tau(P))=f(\tau(P))\psi(n,\tau(P)).
$$
Consequently,
$$
 \overline{L_m\psi(n,\tau(P))}=\bar{L}_m\overline{\psi(n,\tau(P))}=\bar{L}_m\psi(n,P)=\overline{f(\tau(P))} \overline{\psi(n,\tau(P))}.
$$
Hence
$$
 \bar{L}_m\psi(n,P)=f(P)\psi(n,P).
$$
From the uniqueness of the operator corresponding to the meromorphic function $f(P)$ we get
$$
 \bar{L}_m=L_m,
$$
hence, the coefficients of $L_m$ are real.

To prove the first part of the theorem we introduce the following function 
$$
\chi(n,P)=\dfrac{\psi(n+1,P)}{\psi(n,P)}.
$$
From the definition of the Baker--Akhiezer function we obtain that the zero and pole divisor of $\chi$ has the form
\begin{equation}\label{eq7v}
(\chi(n,P))= \gamma(n+1)+ P_{n}-\gamma(n)-q,\quad n\in{\mathbb Z}.
\end{equation}

\begin{lemma}
 Operators from ${\mathcal A}$ have $N$-periodic coefficients if and only if $$\chi(n+N,P)=\chi(n,P).$$
\end{lemma}
{\it Proof of Lemma 1.}
Let us prove the inverse part of the lemma. We assume that the coefficients of all operators from ${\mathcal A}$ are periodic. This means that the operator $T^N$ commutes with all operators from ${\mathcal A}$, i.e.,  $T^N\in{\mathcal A}$. This also means that there is a meromorphic function $\lambda(P)$ on $\Gamma$ with the unique pole in $q$ of order $N$ such that
\begin{equation}
T^N\psi(n,P)=\psi(n+N,P)=\lambda(P)\psi(n,P).
\end{equation}
We have
\begin{equation}
\chi(n+N,P)=\dfrac{\psi(n+1+N,P)}{\psi(n+N,P)}=\dfrac{\lambda(P)\psi(n+1,P)}{\lambda(P)\psi(n,P)}=\chi(n,P).
\end{equation}
Let us prove the direct part of the lemma. We assume that $\chi(n+N,P)=\chi(n,P)$. 
We introduce a rational function on $\Gamma$
$$
 \lambda(P)=\chi(0,P)\dots \chi(N-1,P)=\psi(N,P).
$$
Then we obtain
$$
T^N\psi(n,P)=\psi(n+N,P)=\chi(n+N-1,P)\psi(n+N-1,P)=
$$
$$\chi(n+N-1,P)\chi(n+N-2,P)\psi(n+N-2,P)=\ldots=
$$
$$
\chi(n+N-1,P)\dots \chi(n,P)\psi(n,P)=
$$
$$
 \chi(0,P)\dots \chi(N-1,P)\psi(n,P).
$$
Hence, 
\begin{equation}\label{eq11}
T^N\psi(n,P)=\lambda(P)\psi(n,P). 
\end{equation}
From (\ref{eq11}) it follows that $T^N\in{\mathcal A}$ since $T^N$ and operators from ${\mathcal A}$ have common Baker--Akhiezer 
eigenfunction.  Moreover $\lambda(P)$ has the unique pole of order $N$ in $q$. Lemma 1 is proved.

Now we can finish the proof of Theorem 1. Let us assume that coefficients of the operators are periodic. Then by Lemma 1 the function $\chi$ is periodic and from (\ref{eq7v}) we have 
\begin{equation}\label{eq12v}
 (\chi(n+N,P))=\gamma(n+N+1)+P_{n+N}- \gamma(n+N)-q.
\end{equation}
Hence, comparing the pole divisors of (\ref{eq7v}) and (\ref{eq12v}) we get $\gamma(n)=\gamma(n+N)$,
and after comparing the zero divisors of  (\ref{eq7v}) and (\ref{eq12v})  we get $P_{n+N}=P_n.$

From the proof of Lemma 1 it follows that the function $\lambda(P)=\psi(N,P)$ has an unique pole q of order $N,$ moreover
$$
 (\lambda(P))=\gamma(N)+P_0+\dots+P_{N-1}-\gamma(0)-Nq=P_0+\dots+P_{N-1}-Nq.
$$
Hence the direct part of Theorem 1 is proven.

Let us assume that there is a meromorphic function $\lambda(P)$ such that
$$
 (\lambda  (P))=P_0+\dots+P_{N-1}-Nq,
$$
and $P_{n+N}=P_n$.
We can suppose  that in the neighborhood of $q$ we have the expansion
$$
  \lambda=k^N+O(k^{n-1}).
$$
Then from  (\ref{eq7v}) we have
$$
 (\chi(n+N))=\gamma(n+N+1)+P_{n+N}-\gamma(n+N)-q=
$$
$$
 \gamma(n+1)+P_n-\gamma(n)-q=\chi(n).
$$
Since $\chi(n)=k+O(1)$ in the neighborhood of $q$, we get $\chi(n+N)=\chi(n)$. Hence by Lemma 1  the coefficient of the operators are periodic.
Theorem 1 is proven.

\subsection{Example}

Let us consider the case of elliptic spectral curve $\Gamma$ given by the equation
$$
 w^2=F(z)=z^3+c_2z^2+c_1z+c_0.
 $$
 The degree of the divisor $\gamma(n)$ is 1. Let
 $$
  \gamma(n)=(\alpha_n,\beta_n)\in\Gamma,\quad \beta_n^2=F(\alpha_n).
$$
Commuting operators of orders 2 and 3 have the forms (see \cite{MauMir})
$$
L_2 = (T + U_n)^2 + W_n,
$$
$$
U_n = \frac{\beta_n+\beta_{n+1}}{\alpha_{n+1} - \alpha_n},
\qquad
W_n = -c_2 - \alpha_n - \alpha_{n + 1},
$$
$$
L_3 = T^3 + (U_n + U_{n + 1} + U_{n + 2}) T^2 + (U^2_n + U^2_{n + 1} + U_n U_{n + 1} +
      W_n - \alpha_{n + 2}) T + 
$$
$$
(U_n (U^2_n + W_n - \alpha_n)  +\beta_n).
$$
The function $\chi(n,P)$ has the form
$$
\chi = \frac{w+ \beta_n}{z - \alpha_n} + \frac{\beta_n +\beta_{n+1}}{\alpha_n - \alpha_{n + 1}}.
$$
The point $P_n=(z_n,w_n)\in\Gamma$ has the coordinates
$$
 z_n=\frac{c_1(\alpha_n+\alpha_{n+1})+\alpha_n\alpha_{n+1}(\alpha_n+\alpha_{n+1})+
 2 c_2\alpha_n\alpha_{n+1}+2(c_0+\beta_n\beta_{n+1})}{(\alpha_n-\alpha_{n+1})^2},
$$
$$
	w_n=\frac{\beta_{n+1}(\alpha_n-z_n)+\beta_n(\alpha_{n+1}-z_n)}{\alpha_n-\alpha_{n+1}}.
$$
If $\alpha_{n+N}=\alpha_n, \ \beta_{n+N}=\beta_n$, then
$$
 \gamma(n+N)=(\alpha_{n+N},\beta_{n+N})=(\alpha_n,\beta_n)=\gamma(n),\quad P_{n+N}=P_n,
$$
and the meromorphic function
$$
 \lambda(P)=\chi(0,P)\dots\chi(N-1,P)
$$
satisfies the conditions of Theorem 1.

\bibliographystyle{plain}

Alina Dobrogowska

Institute of Mathematics, University of Bialstok,
 
 Ciolowskiego 1M, 15-245 Bialstok, Poland,

E-mail: alina.dobrogowska@uwb.edu.pl

Andrey Mironov

Sobolev Institute of Mathematics, Novosibirsk, Russia, and

Novosibirsk State University, Novosibirsk, Russia.

E-mail: mironov@math.nsc.ru

\end{document}